\documentclass[twocolumn]{jpsj2}

\def\runtitle{Quasiparticle Excitations outside the Vortex Cores 
in \mg\/ Probed by Muon Spin Rotation}
\def\runauthor{Kazuki \textsc{Ohishi}, \textit{et al}.}
\newcommand{\yn}{YNi$_{2}$B$_{2}$C}

\newcommand{\tc}{$T_{c}$}
\newcommand{\msr}{$\mu$SR}
\newcommand{\la}{$\lambda$}
\newcommand{\mg}{MgB$_2$}

\title{%
Quasiparticle Excitations outside the Vortex Cores in \mg\/ 
Probed by Muon Spin Rotation
}

\author{%
Kazuki \textsc{Ohishi}\thanks{Present address: 
Institute of Materials Structure Science,
High Energy Accelerator Research Organization (KEK), 
Tsukuba, Ibaraki 305-0801, Japan.}, 
Takahiro \textsc{Muranaka}, Jun \textsc{Akimitsu}, \\ 
Akihiro \textsc{Koda}$^1$, Wataru \textsc{Higemoto}$^1$ and Ryosuke
\textsc{Kadono}$^1$\thanks{Also at School of Mathematical and Physical Science,
The Graduate University for Advanced Studies.}
}

\inst{%
Department of Physics, Aoyama-Gakuin University,
Setagaya-ku, Tokyo 157-8572, Japan \\
$^1$Institute of Materials Structure Science,
High Energy Accelerator Research Organization (KEK), 
Tsukuba, Ibaraki 305-0801, Japan
}

\recdate{\today}

\abst{%
The magnetic penetration depth \la\/ in the mixed state of 
\mg\/ has been determined microscopically from the field 
distribution around magnetic vortices probed by
muon spin rotation. Both the temperature and magnetic field 
dependence of \la\/ strongly suggest the presence of 
delocalized quasiparticles around the vortex cores. 
In particular, the effect of Doppler shift has been clearly observed as a 
finite gradient of the field dependence of \la, strongly suggesting an anisotropic
order parameter with the region of a vanishingly small energy gap.}

\kword{%
\mg\/, flux-line lattice state, penetration depth, \msr\/
}

\begin{document}
\sloppy
\maketitle
The revelation of superconductivity in a binary intermetallic compound,
\mg, has attracted much interest because it exhibits a transition
temperature \tc\/ ($\simeq$ 39~K) almost two times higher than those of all
other intermetallic superconductors known to date \cite{Nagamatsu:01}.
The most interesting issue associated with this compound is whether or not it
belongs to the class of conventional Bardeen-Cooper-Schrieffer (BCS)
type superconductors. 
To date, most experimental results have favored a phonon mediated
superconductivity. 
The boron isotope effect \cite{Budko:01}, photoemission spectroscopy 
\cite{Takahashi:01}, $^{11}$B-NMR \cite{Kotegawa:01}, 
Raman spectroscopy \cite{XKChen:01}, 
tunneling measurements \cite{Ekino,Karapetrov:01,Sharoni:01,Schmidt:01}, and 
optical conductivity data \cite{Pronin:01} are all 
consistent with the conventional BCS $s$-wave pairing.
On the other hand, calculations of the band structure and the phonon spectrum 
predict a double energy gap \cite{Kortus:01,Liu:01}, a larger gap attributed 
to two-dimensional $p_{x-y}$ orbitals, and a smaller gap attributed to 
three-dimensional $p_z$ bonding and antibonding orbitals. 
Experimental results of specific heat measurements \cite{Bouquet:01,Wang:01}, 
point-contact spectroscopy \cite{Szabo:01}, photoemission spectroscopy 
\cite{Tsuda:01}, 
scanning tunneling spectroscopy \cite{Giubileo:01} and penetration depth 
measurements \cite{Manzano:01} have supported this scenario.
 
It must be noted that earlier experiments including 
muon spin rotation (\msr) \cite{Panagopoulos:01} 
and ac susceptibility \cite{Panagopoulos:01,XHChen:01} 
performed on polycrystalline samples revealed a quadratic behavior 
of \la\ at low temperatures, from which they inferred the 
presence of line nodes in the order parameter. 
However, the recent \msr\ analysis has demonstrated that 
such behavior can also be explained by assuming the double energy gap
(without resorting to the line nodes) \cite{Niedermayer:02}. 
Thus, the temperature dependence of \la\ provides limited
information for determining the structure of the order parameter.
This situation can be improved by studying the magnetic
field dependence of \la, where the quasiparticle excitation
is controlled by the Doppler shift\cite{Volovik:93} which is independent of
the thermal excitation.
In order to obtain more detailed information on the
order parameter in \mg, we have observed the field dependence of \la\/ 
over a wide range of magnetic field up to 5~T.
 
The magnetic penetration depth \la\/ is determined by the quasiparticle
excitations outside the vortex cores, and thus it provides an excellent
measure of the structure of the order parameter in the 
flux-line lattice (FLL) state.
The \msr\/ technique is a powerful microscopic tool for obtaining the fundamental
length scale such as \la\/ in the bulk type II superconductors.
Implanted muons randomly probe the local magnetic fields induced by the
FLL, yielding the spectral density $n(B)$ which is directly
related to the spatial field distribution $B(r)$. 
While the \msr\ spectra in a single crystalline specimen 
can be compared directly with $n(B)$ calculated from the
spatial field distribution $B(r)$, the spectra in a 
polycrystalline specimen are subject to the modulation
of line shape due to various kinds of inhomogeneity.
Even in such a situation, the time-dependent \msr\ spectra
is approximately described by a Gaussian damping 
$\exp(-\sigma^2t^2/2)$ with $\sigma$ being primarily determined by
the second moment of $n(B)$.  

In this letter, we report on the temperature and field
dependence of \la\ deduced from those of $\sigma$ in polycrystalline
\mg. A special precaution has been taken in selecting the field
for the temperature scan, in order to avoid the effect of random flux pinning
near the lower critical field $H_{c1}$.  
Analysis based on the two-gap model yields 
$\Delta_1$ = 4.9(1)~meV and $\Delta_2$ = 1.2(3)~meV. 
The penetration depth at $T$ = 0~K is estimated to be 103.9(1.0)~nm. 
We also found that \la\ at $T \simeq$ 10~K exhibits a significant 
increase with almost linear dependence on the applied magnetic field,
which can be understood by considering the Doppler shift of the quasiparticle 
excitation associated with the anisotropic order parameter\cite{Volovik:93}.  
However, the gradient against the field is considerably
small compared with that in $d$-wave superconductors.
These results indicate the existence of 
excess quasiparticle excitations outside the vortex cores in \mg, 
strongly suggesting that there is an anisotropic structure in the order parameter
with a nodal region smaller than that for the $d$-wave pairing. 

The polycrystalline sample of \mg\/ used in this experiment
had a surface area of $\sim$ 50~mm$^{2}$. 
The superconducting transition temperature $T_{c}$ determined
from resistivity and susceptibility measurements was 38.5~K \cite{Nagamatsu:01}.
\msr\/ experiment was performed on the M15 beamline at TRIUMF
which provides a muon beam with the momentum of 29~MeV/c.
A muon-veto counter system was adopted to eliminate
positron events from muons which missed the sample so that
the relative yield of such events was less than 5\%\/ of the
total positron events. An experimental setup
with a high time resolution was employed to measure the transverse field (TF-)
\msr\/ time spectra up to 5~T.
The sample was field cooled at the measured
magnetic fields in order to eliminate the effect of flux pinning.

Since the muons stop randomly on the length scale of the
FLL, the muon spin precession signal $\hat{P}(t)$ provides
a random sampling of the internal field distribution $B(r)$,
\begin{align}
\hat{P}(t) \equiv P_x(t)+iP_y(t)&=\int_{-\infty}^\infty
n(B)\exp(i\gamma_\mu Bt)dB,\\
n(B)&=\frac{dr}{dB},
\end{align}
where $\gamma_\mu$ is the muon gyromagnetic ratio (= $2\pi\times$135.53~MHz/T),
and $n(B)$ is the spectral density determined by the local field distribution.
These equations indicate that the real amplitude of the Fourier transformed
muon precession signal corresponds to the spectral density $n(B)$.
The London penetration depth in the FLL state is
related to the second moment $\langle (\Delta B)^2\rangle$
of $n(B)$ reflected in the \msr\/ line shape\cite{Brandt:88}.
For polycrystalline samples,
a Gaussian distribution of local fields is a good approximation, where
\begin{align}
\hat{P}(t) &\simeq\exp(-\sigma^2t^2/2)\exp(i\gamma_\mu Ht)\\
\sigma&=\gamma_\mu\sqrt{\langle (\Delta B)^2\rangle },
\end{align}
with $H$ being the external magnetic field. 
For the case of an ideal triangular FLL with the isotropic effective
carrier mass $m^*$ and a cutoff $K\approx1.4/\xi_v$ provided by the
numerical solution of the Ginsburg-Landau theory, 
the London penetration depth $\lambda$ can be deduced from $\sigma$ 
using the following equation\cite{Brandt:88}:
\begin{equation}
\sigma\ [\mu{\rm s^{-1}}] = 4.83\times 10^4 
(1-h)[1+3.9(1-h)^2]^{1/2} \lambda^{-2}\ [{\rm nm}],
\label{Brandt}
\end{equation}
where $h=H/H_{c2}$. It should be noted that eq.~(\ref{Brandt}) provides 
the field dependence of $\sigma$ when \la\/ is a constant. 
Here, $\lambda$ is related to the superconducting carrier density
$n_s$ as follows:
\begin{equation}
\lambda^2=\frac{m^*c^2}{4\pi n_se^2},\label{lambda}
\end{equation}
indicating that $\lambda$ is enhanced upon the reduction of $n_s$
due to the quasiparticle excitations.

\begin{figure}[t]
\begin{center}
\rotatebox[origin=c]{0}{\includegraphics[width=0.45\textwidth]{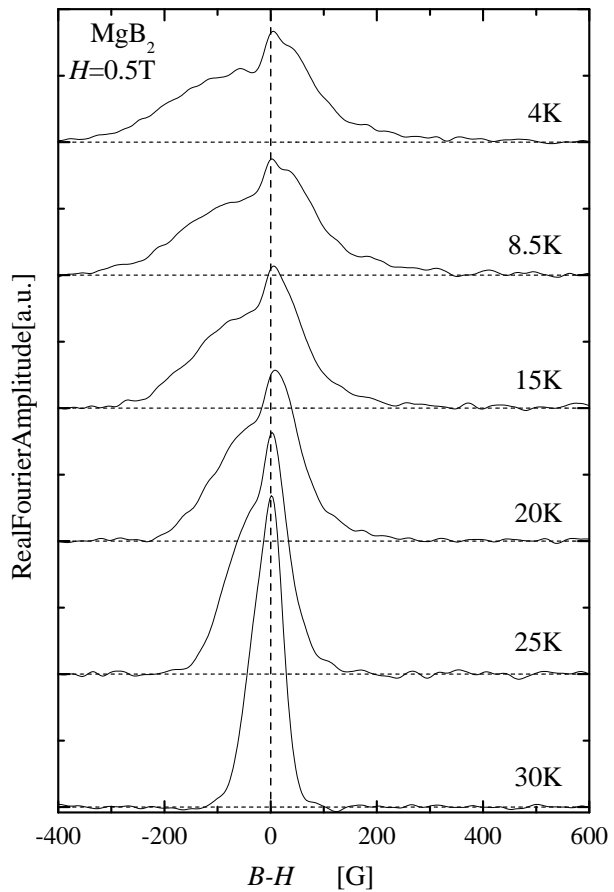}}\\
\end{center}
\caption{FFT spectra of \msr\ signal in \mg\ at $H \simeq$ 0.5 T 
under several temperatures.}
\label{FFT}
\end{figure}

Figure \ref{FFT} shows the fast Fourier transforms (FFT) of the
muon precession signal in \mg\ for different temperatures at $H \simeq$ 0.5~T,
where the real amplitude of FFT corresponds to $n(B)$ in the FLL state 
convoluted with additional damping
due to small nuclear dipolar fields. As is most explicit in the FFT spectrum at
$T$~=~8.5~K, the line shape is characterized by a broad peak near $B - H \sim 0$
and a satellite at lower field (with a small background peak right at $B = H$).
This is a typical feature often seen in polycrystalline powder samples with 
a lower Meissner fraction, where the sample consists of both superconducting 
and normal domains with the typical size of a few microns ($\geq$ \la).
The magnetic field distribution at the normal domain is shifted to a high
field because of the demagnetization associated with the Meissner effect
in superconducting domains. Thus, the peak at the lower field corresponds to 
the signal from the superconducting domains.

Considering the double peak structure in the FFT spectra in Fig.~\ref{FFT},
we adopted two components with the following empirical form for analyzing 
the data in the time domain:
\begin{align}
A_0\hat{P}(t)&=\sum_{j=1,2}A_j(t)\exp[i(\gamma_\mu B_jt+\phi)],\\
A_j(t)&=A_j\exp\left[-(\sigma_j t)^2/2\right],
\end{align}
where $A_0$ is the total positron decay asymmetry,
$A_j$ ($j=1,2$) the partial asymmetry for respective
components, $B_j$ the central frequencies, $\phi$ the initial phase,
and $\sigma_j$ the muon depolarization rates.

\begin{figure}[t]
\begin{center}
\rotatebox[origin=c]{0}{\includegraphics[width=0.45\textwidth]{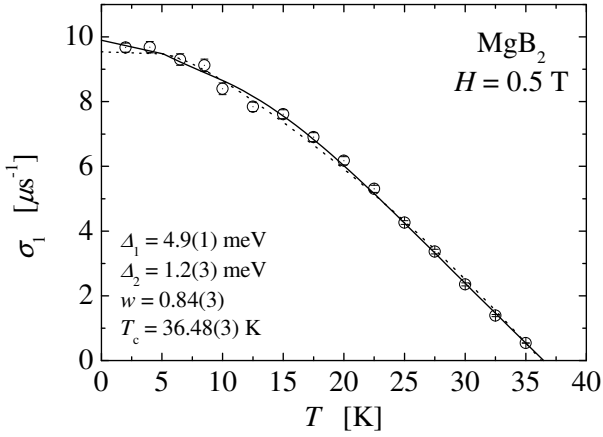}}\\
\end{center}
\caption{Temperature dependence of the muon depolarization rate $\sigma_1$ 
under $H$ = 0.5 T. The solid curve is the result of fitting by eq. (\ref{2gap}).}
\label{T-dep}
\end{figure}

We show the temperature dependence of the muon depolarization rate $\sigma_1$
(which corresponds to the linewidth of the lower frequency peak
in Fig.~\ref{FFT}) in Fig.~\ref{T-dep}.
These data were obtained under a field $H \simeq$ 0.5~T which is well above 
$H_{c1}\sim10^{-2}$ T. 
Following the results of specific heat \cite{Bouquet:01,Wang:01}, 
Raman \cite{XKChen:01} and \msr\/ measurements 
\cite{Niedermayer:02}, the data were fitted by the 
two-gap model \cite{Bouquet:01-2} which is described as follows: 
\begin{eqnarray}
\sigma (T)=\sigma (0)-w\cdot\delta\sigma (\Delta_1,T) - 
(1-w)\cdot\delta\sigma (\Delta_2,T), \label{2gap}\\
\delta\sigma (\Delta,T)=\frac{2\sigma (0)}{k_BT}\int^\infty_0f(\varepsilon,T)\cdot
[1-f(\varepsilon,T)]d\varepsilon, \\
f(\varepsilon,T)=\left(1+e^{\sqrt{\varepsilon^2+\Delta(T)^2}/k_BT}\right)^{-1},
\end{eqnarray}
where $w$ is the ratio of gap energy between the two gaps, 
$k_B$ the Boltzmann constant, $f(\varepsilon,T)$ 
the Fermi distribution of quasiparticles, and $\Delta(T)$ 
the BCS gap energy \cite{Muhlschlegel:59}. 
The solid line in Fig.~\ref{T-dep} is the best fit 
result with $\Delta_1$ = 4.9(1)~meV, 
$\Delta_2$ = 1.2(3)~meV, $w$ = 0.84(3) and $T_c$ = 36.48(3)~K. 
The dotted line shows the result of fitting using the values of $\Delta_1$, 
$\Delta_2$ and $w$ in ref. \citen{Niedermayer:02}. 
Although our result shows reasonable agreement with  
the earlier one \cite{Niedermayer:02}, the value of $\Delta_2$
is considerably smaller than the reported value of 2.6(2)~meV.

\begin{figure}[t]
\begin{center}
\rotatebox[origin=c]{0}{\includegraphics[width=0.45\textwidth]{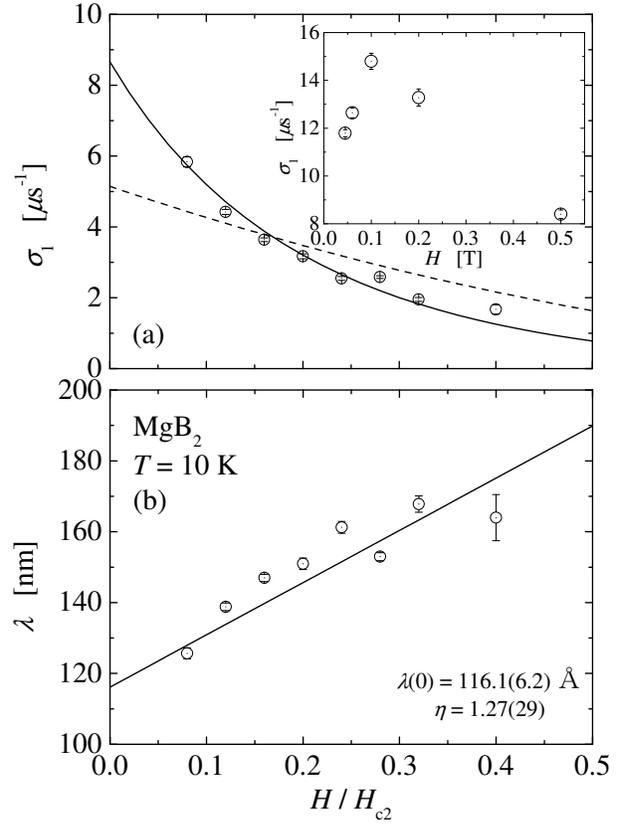}}\\
\end{center}
\caption{Magnetic field dependence of (a) the muon depolarization rate $\sigma_1$
and (b) the magnetic penetration depth \la\/ at $T \simeq$ 10 K.
Dashed and solid lines are the fittings using (a) eq. (\ref{Brandt}) with
 a constant \la\/ or \la\/ proportional to \la\/(0)$[1+\eta\cdot h]$ and 
(b) eq.~(\ref{eta}). $H_{c2}$ at this temperature is approximately 12.5~T.}
\label{H-dep}
\end{figure}

Figure~\ref{H-dep}(a) shows the magnetic field 
dependence of the muon spin relaxation rate 
$\sigma_1$ with an inset showing the low 
field region ($H < 0.6$~T). 
As seen in the earlier report\cite{Niedermayer:02}, 
the effect of random flux pinning is observed 
as a peak of $\sigma_1$ in the low field region (see inset of
Fig.~\ref{H-dep}(a)). 
However, $\sigma_1$ decreases 
with increasing field above $H$ = 0.1~T, 
indicating that the distortion of FLL is reduced by
increasing inter-vortex interaction and that the
depolarization is predominantly determined by the
intrinsic $n(B)$. 
The dashed line in Fig.~\ref{H-dep}(a) shows 
the fitting result by eq.~(\ref{Brandt}) with $H_{c2}$ = 12.5~T as 
determined by resistivity measurements \cite{Muller:02}. 
(Here, we discuss the magnetic field dependence of $\sigma_1$ for 
the data above $H$ = 0.5~T to avoid the remnant effect of
flux pinning at lower fields.) 
Compared with the field dependence of $\sigma_1$, 
the dashed line does not reproduce the data, 
indicating that \la\/ is not a constant but it increases
with increasing external field.

The field dependence of \la\/ estimated from eq.~(\ref{Brandt}) 
is shown in Fig.~\ref{H-dep}(b). 
It clearly exhibits a strong field dependence where \la($h$) 
increases almost linearly with $h$. 
This is similar to the cases of \yn\/ \cite{Ohishi:02}, 
NbSe$_2$ \cite{Sonier:97} and
high-\tc\/ cuprate superconductors \cite{Sonier:00}, where the increase 
of \la\/ is attributed
to the anisotropic order parameters and the associated
nonlinear effect due to the Doppler shift of the quasiparticles 
in the nodal region ($\Delta(k)\simeq0$) \cite{Volovik:93}. 
The field dependence of \la\/ is expected to be stronger 
when the phase space satisfying $\Delta(k)\simeq0$ has larger
volume\cite{Amin:00}. A fitting by the relation
\begin{equation}
\lambda(h) = \lambda(0)[1+\eta\cdot h]
\label{eta}
\end{equation}
provides a dimensionless parameter $\eta$ which represents 
the strength of the pair breaking effect. 
We obtain $\eta$ = 1.27(29) with $\lambda(0)$ = 116.1(6.2)~nm 
which is shown as the solid line in Fig.~\ref{H-dep}(b). 
The obtained value of $\eta$ is intermediate between that in \yn\/ 
and NbSe$_2$ (e.g., $\eta$ = 0.97 
at 0.2\tc\/ \cite{Ohishi:02}, $\eta$ = 1.61 at 0.33\tc\/
\cite{Sonier:97}, respectively) and is smaller than those in 
$d$-wave superconductors (e.g., $\eta$ = 5.5 $\sim$ 6.6 
for cuprates \cite{Sonier:00}). 
The solid line in Fig.~\ref{H-dep}(a) is the result of fitting with 
the relation of eq.~(\ref{Brandt}), with \la\/ represented by eq.~(\ref{eta}).

Our result on the temperature dependence of \la\/ is qualitatively
consistent with earlier results \cite{Niedermayer:02}, 
suggesting that the order parameter in 
\mg\/ may be effectively described by adopting the two-gap
model. However, the observed  field dependence of \la\ is not
expected for the isotropic order parameter irrespective of the multiplicity of
the band structure and the associated gap energy.  
Considering that the current result on the field dependence of \la\ 
was obtained at $T\simeq$ 10~K, this energy scale of 
$\varepsilon\equiv k_BT\sim1$ meV
places an upper boundary on the smaller gap energy $\Delta_2$ in order
to explain the observed effect of the Doppler shift.
Since our estimation for $\Delta_2=1.2(3)$~meV is very 
close to $\varepsilon$, the observed $H$-linear 
behavior of \la\/ may be attributed to the quasiparticle excitations 
in the vicinity of the smaller gap. While this cannot be distinguished 
from the case of a nodal structure in the order parameter 
(i.e., considering a region where $\Delta(k)\ll\varepsilon$), 
our result is clearly inconsistent with the two-gap model with 
$\Delta_2\gg\varepsilon$. 
On the other hand, it is also quite unlikely that 
the $d$-wave pairing is realized in \mg, because the coefficient $\eta$  
is much smaller than those in high-\tc\/ cuprates. 
The recent observation that the order parameter in \yn\/ 
(where the pairing symmetry has been identified as $s$-wave \cite{Shu:96})
has point nodes \cite{Izawa:02} exhibits 
a good correspondence with the intermediate value of $\eta (\sim 1)$ obtained by 
\msr \cite{Ohishi:02}, suggesting that there is a similar situation in \mg.
Thus, the present \msr\/ result leads us to conclude that the order parameter
in \mg\ has a structure with an energy gap smaller than 
$\varepsilon\simeq1$ meV. The field dependence of \la\/ measured
at a much lower temperature would provide more useful information
to distinguish the anisotropic order parameter from the isotropic one described by
the two-gap model.

In summary, we have performed TF-\msr\/ measurements in \mg\/ to obtain the 
temperature and magnetic field dependence of the penetration depth \la\
and the associated spin relaxation rate $\sigma_1$. 
Our result is perfectly in line with the presence of an anisotropic order parameter
with a nodal structure, and it sets an upper boundary
$\varepsilon\simeq1$~meV for the smaller gap energy in the two-gap model.
The magnetic field dependence of \la\/ exhibits a 
linear dependence on the external field up to 5~T with the gradient 
$\eta$ being considerably smaller than that in $d$-wave superconductors, 
which may disfavor the occurrence of $d$-wave pairing in \mg.

We thank J.-M. Poutissou of TRIUMF for the allocation of beamtime and 
all the TRIUMF \msr\/ staff for their technical support. 
This work was partially supported by a JSPS Research Fellowships for
Young Scientists, by a Grant-in-Aid for Scientific Research on Priority
Areas from the Ministry of Education, Culture, Sports, Science, and
Technology, Japan, and also by a Grant from CREST, JST, Japan.

\end{document}